\documentclass[apjl]{emulateapj}

\usepackage{amsmath}
\usepackage{wasysym}
\usepackage[dvipsnames]{xcolor}
\usepackage{multirow}
\usepackage[caption=false]{subfig}
\usepackage[colorlinks,citecolor=blue]{hyperref}

\usepackage{times}
\shorttitle{New limits on thermally annihilating Dark Matter from Neutrino Telescopes}
\shortauthors{Lopes et al.}

\begin{document}


\title{New limits on thermally annihilating Dark Matter from Neutrino Telescopes}


\author{J. Lopes}
\affil{CENTRA, Physics Department, Instituto Superior T\'ecnico, Universidade de Lisboa, Portugal}
\email{josevlopes@ist.utl.pt}

\and

\author{I. Lopes}
\affil{CENTRA, Physics Department, Instituto Superior T\'ecnico, Universidade de Lisboa, Portugal}



\begin{abstract}
We used a consistent and robust solar model to obtain upper limits placed by neutrino telescopes, such as IceCube and Super-Kamiokande,  on the Dark
Matter-nucleon scattering cross-section, for a general model of Dark Matter with a velocity 
dependent ($p$-wave) thermally
averaged cross-section. In this picture, the Boltzmann equation for the Dark Matter abundance is numerically solved
satisfying the Dark Matter density measured from the Cosmic Microwave Background (CMB). We show that for lower
cross-sections and higher masses, the Dark Matter annihilation rate drops sharply, resulting in upper bounds  on the
scattering cross-section one order of magnitude above those derived from a velocity independent ($s$-wave) annihilation
cross-section. Our results show that upper limits on the scattering cross-section obtained from Dark Matter annihilating 
in the Sun are sensible to the uncertainty in current standard solar models, fluctuating a maximum of 20 \% depending on 
the annihilation channel.
\end{abstract}


\section{Introduction}
The joint efforts of physicists in the last four decades has resulted in solid evidence, not only at astrophysical scales, but also cosmological, which leave no doubt that our Universe is mainly populated by a still undetected non-interactive type of
matter, the so-called Dark Matter, which nature is still unknown. Amongst the numerous theories devised to solve this problem, the picture of a weakly interactive massive particle (WIMP) arises as the most favourable, since the dark matter abundance inferred today from the Cosmic Microwave Background (CMB) matches the abundance of a relic particle with an annihilation cross-section of the order of the weak scale. Furthermore, new particle physics theories motivated by different reasons, provide natural candidates for this type of matter, making this a highly interdisciplinary field of investigation. \\ 

In the picture of particle Dark Matter, WIMPs that populate the Milky Way can be gravitationally captured by the Sun~\citep{steigman78}. Since WIMPs are mandatorily stable, they will accumulate inside Sun and annihilate to become standard model
particles. This annihilation will produce a distinctive neutrino signal that can be detected in current and projected neutrino detectors, providing an excellent indirect survey to Dark Matter properties, which has been extensively studied~\citep{silk85,gaisser86,griest87,wilkstrom09}.\\

However, most indirect Dark Matter searches focus on simple models where WIMPs are Majorana particles and annihilate through an $s$-wave, velocity independent thermally averaged cross-section. 
In these models, 
it is usual to fix $\langle \sigma v \rangle \sim 10^{-26} \ \text{cm}^3 \text{s}^{-1}$ , in order to respect the Dark Matter density~\citep{kolb89}, which is precisely determined from CMB measurements ~\citep{planck15}. Theoretically, the Dark Matter abundance is defined by the time of WIMP freeze out, which happens when the universe's temperature drops below the WIMPs mass, resulting in the decoupling of Dark Matter particles from the primordial universe thermal bath. This will result in a constant number of WIMPs per co-moving volume which corresponds to the one measured today. \\

Despite the fact that the usual approach in literature is to use a constant thermally averaged cross-section, there are a large number of well-motivated models which have a $p$-wave contribution, (i.e. a dependence in the relative velocity between WIMPs) to the annihilation cross-section which can be dominant. Models where Dark Matter is a Majorana particle, such as the \textit{neutralino}, a natural candidate which arises in the Minimal Super symmetric Standard Model (MSSM), $s$-wave annihilation to fermion anti-fermion pairs is helicity suppressed by a factor of $(m_f/m_{\chi})^2$,
where $m_{\chi}$ is the WIMPs mass~\citep{sheldon10,goldberg83}. Furthermore, due to $CP$ conservation, final states with $CP$=+1, are only accessible through $p$-wave annihilations ($s$-wave states for two identical majorana fermions are $CP$=-1). Hence, neutralino annihilation to $HH$ or any combination of the vectorial bosons $W$ and $Z$, can only occur through $p$-wave annihilation ~\citep{drees92}. Another well known example is the case of parity conserving minimal extensions to the Standard Model (SM) with a fermionic Dark matter candidate - a gauge singlet Dirac fermion - in which annihilation to scalar states with even parity, such as $\chi \chi \rightarrow HH$, does not receive a contribution from the $s$-wave annihilations ~\citep{kim07}. \\

In this article we obtain new limits on the Dark Matter scattering cross-section from the upper limits on the neutrino fluxes measured by the Super-Kamiokande and IceCube neutrino telescopes, using a general model where $p$-wave annihilation is the leading contribution to the total annihilation cross-section, i.e., where we focus in the second term in the thermally averaged annihilation cross-section expansion, \\

\begin{equation}
\left \langle \sigma v \right \rangle = a + b \left \langle v^2 \right \rangle + \mathcal{O}(\langle v^4 \rangle ) \simeq \frac{b'}{x},
\label{eq:annCS}
\end{equation}

with $x = m_{\chi}/T$, and where we also assumed that the $s$-wave contribution to the annihilation, as well as higher order terms with $ \mathcal{O}\left( \langle v^4 \rangle \right)$, are negligible. 

The coefficient $b$ in \ref{eq:annCS} is assumed constant and obtained taking into account the Dark Matter density at the time of freeze-out, which is roughly the same as today (see sec. \ref{sec:DMabundance}).

It is important to note that, since we are mainly interested in the epochs from the moment of WIMP freeze-out, the expansion in \ref{eq:annCS} is only accurate if WIMPs freeze out at non-relativistic velocities, i.e. if $T_F < m_{\chi}$, where $T_F$ is the temperature of freeze-out. However, in most cases this happens at $T_F \simeq m_{\chi}/20 \ll m_{\chi}$ ~\citep{jungman96}, which means that the thermally averaged annihilation cross-section in \ref{eq:annCS} is a reliable approximation for our analysis.\\ 

In section \ref{sec:DMabundance} we obtain the coefficient $b$ by solving the Boltzmann equation for a relic particle with a velocity dependent annihilation cross-section. In sec. \ref{sec:DMsun} we present the formalism needed to compute the neutrino flux from dark matter annihilation in this picture, as well as the stellar evolution code used to compute the Dark matter capture and annihilation to SM particles. In sec. \ref{sec:DMresults} we present our results followed by some final remarks.   

\section{Velocity dependent relic abundance}
\label{sec:DMabundance}

The Dark Matter particle co-moving number density $n_{\chi}$ is governed the Boltzmann equation ~\citep{kolb89}
\begin{equation}
\frac{dn_{\chi}}{dt} = - 3Hn_{\chi} -\langle \sigma v \rangle (n_{\chi}^2-n_{{\chi},\text{eq}}^2),
\label{eq:boltzmann}
\end{equation}
where $H$ is the Hubble parameter, and $n_{{\chi},\text{eq}}$ is the WIMP's number density when in equilibrium with the thermal plasma. Equation \ref{eq:boltzmann} takes into account the Universe's expansion rate as well as the annihilation and production of WIMPs from the thermal plasma (first and second terms in the r.h.s. of eq. \ref{eq:boltzmann}). To simplify equation \ref{eq:boltzmann} it is usual to use the law of entropy conservation defining the number density $n_{eq}$ in terms of the universe's total entropy $Y\equiv n_{\chi}/s$, and changing the independent variable from $t$ to  $x\equiv m_{\chi}/T$, where $T$ is the photon temperature. Using the annihilation cross-section given in \ref{eq:annCS}, where $\langle \sigma v \rangle \propto x^{-1}$, the Boltzmann equation eq. \ref{eq:boltzmann} can be written as 
\begin{equation}
\frac{dY}{dx}=- \Lambda x^{-3} (Y^2-Y_{\text{eq}}^2),
\label{eq:simpBoltz}
\end{equation}
with 
\begin{equation}
\Lambda = \sqrt{\frac{\pi}{45 g_{\rho}}}g_{s}m_{Pl}m_{\chi} b',
\label{eq:lambda}
\end{equation}
where $m_{Pl} = $ is the planck's mass. The effective number of relativistic degrees of freedom in the Universe contributing to the energy density, $g_\rho$, can be approximated by a step-function ~\citep{dent10} and in the epoch of WIMP freeze-out is essentially the same as the effective number of relativistic degrees of freedom contributing to the total entropy, $g_s$. For this reason, hereafter we will use $g \equiv g_s \equiv g_\rho$.

In this article we are interested in the standard case where WIMPs fully thermalize with the thermal plasma in the early Universe. In this case, in the early Universe (for low $x$), $Y$ tracks the equilibrium number density $Y_{\text{eq}}$ since WIMPs are constantly annihilating and being produced. When the annihilation rate drops below the expansion rate, WIMPs freeze-out, i.e. they fall out of thermodynamic equilibrium and their abundance is fixed. Again, assuming that WIMPs usually freeze-out at temperatures below their mass $T_F \ll m_{\chi}$, one can use a non-relativistic Maxwell-Boltzmann distribution for $Y_{\text{eq}}$ for a WIMP with $g_{\chi}$ degrees of freedom,

\begin{equation}
Y_{\text{eq}}(x)= \frac{45}{2 \pi^4}\sqrt{\frac{\pi}{8}} \frac{g_{\chi}}{g}x^{3/2}e^{-x}.
\label{eq:yeq}
\end{equation}

\begin{figure}[!htb]
	\centerline{
		\includegraphics[width=\columnwidth]{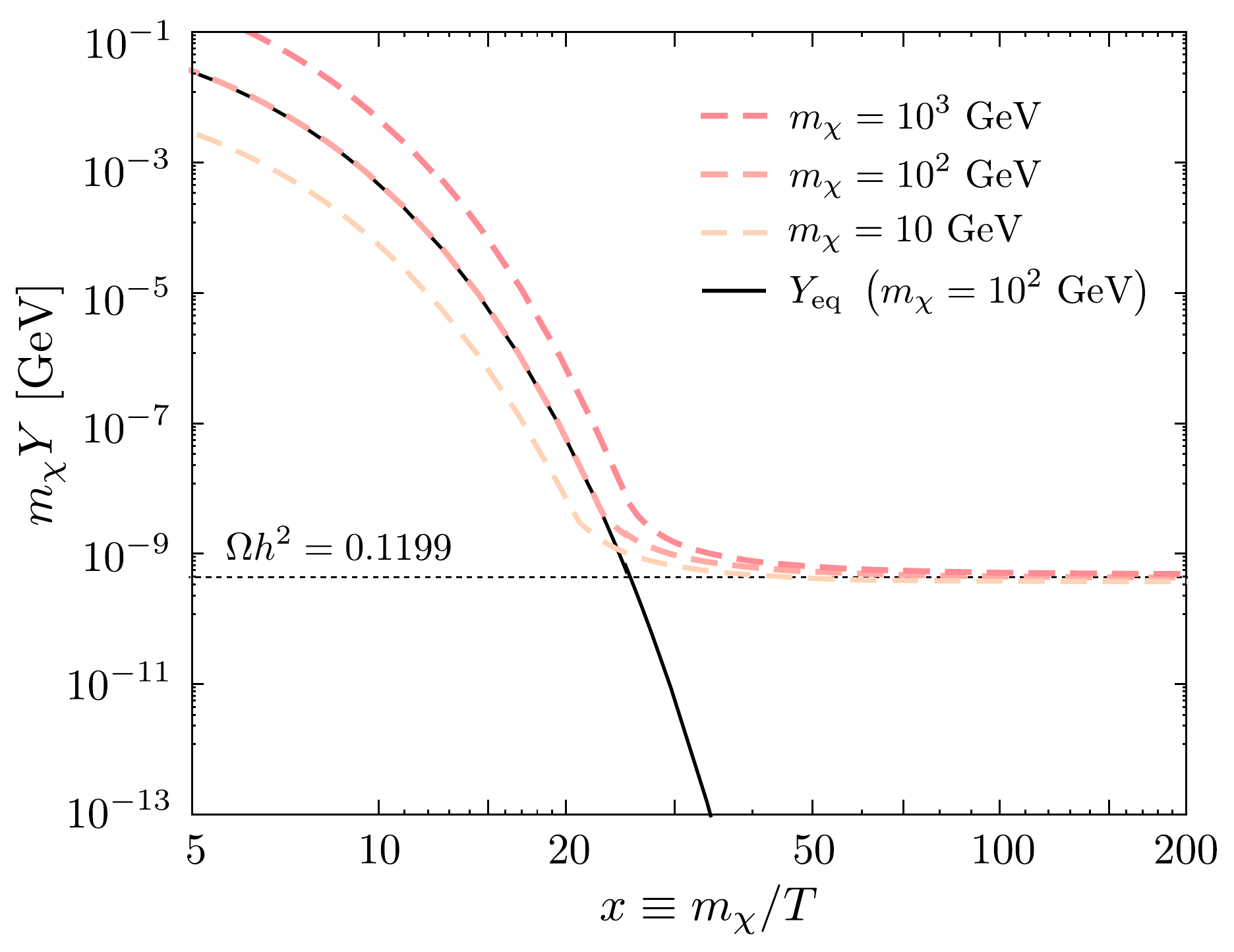}}
	\caption{WIMP number density evolution with $\langle \sigma v \rangle = 9.60 \times 10^{-25} x^{-1}\ \text{cm}^3 \ \text{s}^{-1}$ solved numerically for different WIMP masses. For the sake of simplicity it is only shown the equilibrium abundance for $m_{\chi}=100 \ \text{GeV}$. The number density corresponding to the WIMP relic density (eq.~\ref{eq:relicDens}) inferred from CMB measurements is also shown (dotted line).}
	\label{fig:boltzmann}
\end{figure}

Numerically solving equation \ref{eq:simpBoltz} with \ref{eq:lambda} and \ref{eq:yeq}, using the condition that $Y=Y_{\text{eq}}$ at $x=1$ we obtained the WIMP number density today, $Y_0$ (fig. \ref{fig:boltzmann}). 
Finally, we can compute the WIMP relic density using
\begin{equation}
\Omega_{\chi}h^2 \equiv \frac{\rho_{\chi}}{\rho_{\text{critical}}} = 2,74\times 10^{8} m_{\chi} Y_{0},
\label{eq:relicDens}
\end{equation}
from where we obtained the value $b'$ which best reproduces the relic abundance measured from the CMB, $\Omega_{\chi}h^2 = 0,119$ ~\citep{planck15}, for a wimp of $m_{\chi} = 100$ GeV is 
\begin{equation}
b' = 9,60 \times 10^{-25} \ \text{cm}^3 \text{s}^{-1}.
\end{equation}

It is important to stress that this result is mainly independent from the mass, as we can see in fig. \ref{fig:boltzmann}, except for logarithmic corrections. We can also see by fig. \ref{fig:boltzmann} that WIMPs freeze-out at a temperature of roughly $x_F \gtrsim 20$ which is consistent with the hypothesis that freeze-out occurs when WIMPs are non-relativistic.\\

\section{Dark Matter Annihilation in the Sun}

\label{sec:DMsun}

Dark Matter particles in the Milky Way Halo can be gravitationally trapped inside the Sun. This capture process will continue until equilibrium with annihilation is attained, which will fix the number of WIMPs inside the Sun. The hot plasma will have impact, not only in the Dark Matter distribution inside the Sun, but also on the neutrino flux resulting from its annihilation.

\subsection{Number of WIMPs in the Sun}

\label{sec:DMsunNumber}

The number of WIMPs in the Sun is governed by three distinct processes: capture, annihilation and evaporation (which is the inverse reaction of capture). When the evaporation process is negligible, which is our case since we are studying WIMPs with $m_\chi > 4$ GeV ~\citep{busoni13}, the number of WIMPs is given by

\begin{equation}
\frac{\text{d}N_{\chi}}{\text{d}t} = C_{\astrosun} - A_{\astrosun}N_{\chi}^2
\label{eq:wimpNumber}
\end{equation}
where $C_{\astrosun}$ is the capture rate, and 
\begin{equation}
A_{\astrosun} \equiv \frac{\int_{\astrosun} \langle \sigma v \rangle n_{\chi}(r)^2 dr^3}{\left[ \int_{\astrosun} n_{\chi}(r) dr^3 \right]^2}
\end{equation}
is the annihilation coefficient, which is integrated over the Sun's volume taking into account the WIMPs distribution, $n_{\chi}(r)$. Equation \ref{eq:wimpNumber} has a straightforward solution, which yields 

\begin{equation}
N_{\chi}(t)= \sqrt{\frac{C_{\astrosun}}{A_{\astrosun}}} \ \text{tanh}(\sqrt{C_{\astrosun}A_{\astrosun}} t).
\label{eq:wimpNumber2}
\end{equation}

Defining the equilibrium time-scale $t_{\text{eq}} = \left ( C_{\astrosun}A_{\astrosun} \right )^{-\frac{1}{2}} $, which represents the time needed to achieve equilibrium between WIMP capture and annihilation, the number of WIMPs in the Sun today is in equilibrium if $t_{\text{eq}} \ll t_{\astrosun}$, where $t_{\astrosun}=4.6$ Gyr. In that case, equation \ref{eq:wimpNumber2} simplifies to
\begin{equation}
N_{\chi}(t_{\astrosun}) \simeq \sqrt{\frac{C_{\astrosun}}{A_{\astrosun}}}
\end{equation}
and the annihilation rate is simply given by
\begin{equation}
\Gamma_A = \frac{1}{2}N_{\chi}^2 A_{\astrosun} \simeq \frac{1}{2}C_{\astrosun}.
\label{eq:annihilationRate}
\end{equation}
As we can see by equation \ref{eq:annihilationRate}, if the WIMP number equilibrium has been attained, the annihilation rate, $\Gamma_A$, and consequently the neutrino flux, will depend exclusively on the Capture Rate, which means that no conclusions can be made regarding the nature of the Dark Matter annihilation cross-section using data from neutrino experiments.\\

In the usual approach, where WIMPs annihilate through $s$-wave, the number of WIMPs in the Sun is in the equilibrium regime for most of the WIMP parameter region of interest, and the last equality of eq. \ref{eq:annihilationRate} is an excellent approximation. However, in our case, where the thermally averaged annihilation cross-section is given by \ref{eq:annCS}, the annihilation coefficient is
\begin{equation}
A_{\astrosun} = \frac{1}{N_{\chi}^2} \frac{b'}{m_{\chi}} \int_{\astrosun} T_{\chi}(r) n_{\chi}(r)^2 dr^3,
\label{eq:annInt}
\end{equation}
which will generally yield $t_{\text{eq}}\gtrsim t_{\astrosun}$, rendering the approximation in \ref{eq:annihilationRate} unreasonable.
For example, 
simple order of magnitude estimations yield that WIMPs with $m_\chi \gtrsim 5$ GeV will mainly populate the inner 6 \% of the Solar radius~\citep[see eq. 2.3 of][]{spergel85}.
, where the temperature is approximately $1.5\times 10^7$ K, which is at least 4 orders of magnitude lower than the freeze-out temperature for a WIMP with the same characteristics. This means that $\langle \sigma v \rangle(T)$ for WIMPs annihilating in the Sun will be dramatically lower than $\langle \sigma v \rangle (T_F)$ at freeze-out, which will in turn (eq.~\ref{eq:annInt}) increase the equilibrium time-scale. This will be further explained in sec.~\ref{sec:resultsEquilibrium}.

\subsection{WIMP distribution and temperature}

\label{sec:DMsunDist}

To compute the annihilation coefficient in eq. \ref{eq:annInt} one needs to know the properties of Dark Matter inside the Sun, namely its distribution and temperature. The WIMPs distribution in the Sun is governed by the Knudsen number,
\begin{equation}
K(t) = \frac{l(0,t)}{r_{\chi}(t)}
\label{eq:knudsen}
\end{equation}
which is the ratio between the WIMPs mean free path in the center of the Sun, and the length-scale of its distribution ~\citep{spergel85},
\begin{equation}
r_{\chi}(t) = \left(\frac{9}{4 \pi}\ \frac{kT_{\text{c}}(t)}{G \rho_{\text{c}}(t)m_{\chi}} \right)^{\frac{1}{2}}
\label{eq:dmLenght}
\end{equation}
where $T_{\text{c}}(t)$ and $\rho_{\text{c}}(t)$ are the solar central temperature and density respectively. If $K\ll 1$, WIMPs are in Local Thermodynamic Equilibrium (LTE) with the solar plasma and reflect its temperature, i.e. $T_{\chi}(r)=T_{\astrosun}(r)$ ~\citep{gould90}. If $K\gtrsim 1$, WIMPs will transport energy non-locally and will have an isothermal distribution characterized by a unique temperature, $T_\chi$ ~\citep{spergel85}. To obtain the WIMP distribution for arbitrary $K(t)$, it is usual to interpolate between the LTE and isothermal distributions. The interpolation is computed using a suppression factor, function of $K(t)$, which accounts for the departure of the LTE regime. WIMPs inside the Sun will also contribute as an effective mechanism of energy transport ~\citep{gould90}, which efficiency will depend on the Knudsen number, $K(t)$, and will be different for the two regimes discussed above, LTE and non-local. This process can have impact in the Sun's structure ~\citep{lopes10,lopes10b,lopes14}, however, in our case this impact is negligible, hence we will not pursue this subject further.

\subsection{The Solar Model}

\label{sec:DMmodel}

In our work we used a stellar evolution code which is based on CESAM ~\citep{morel97} that has been further developed to include not only Dark Matter capture and annihilation, but also its energy transport and distribution inside the Sun ~\citep{lopes11}, as described in sec. \ref{sec:DMsunNumber} and \ref{sec:DMsunDist}. The reference model without Dark Matter is calibrated to achieve a Standard Solar Model (SSM) ~\citep{turck93} in full agreement with the predominant solar models used in the literature ~\citep{bahcall05,serenelli09}.


A brief remark regarding the theoretical uncertainty in current Solar models, which is mainly caused by the uncertainty on the heavy element abundances~\citep{bahcall06}, should be made at this point. Recent independent determinations of the photospheric heavy element abundances~\citep{asplund09,caffau11} have led to lower metallicity to Hydrogen ratio when compared with older estimations by~\citep{grevesse98}. Despite using complex and improved solar atmosphere models, these downward revisions of the photospheric heavy element abundances result in Solar models hard to reconcile with Helioseismological and Solar neutrino data~\citep{basu04,bahcall04,turck04}, in opposition to the older estimation by~\citet{grevesse98} which is in agreement with observations. The solution to this discrepancy, known as the Solar abundance problem, is yet to be found, as well as the consensus on which heavy element mixture best reproduces the actual Sun \citep[see][and ref. therein]{haxton13}. For this reason we decided to use both the heavy element mixture by \citep[][ AGSS09]{asplund09} with a metallicity to Hydrogen ratio of $\left(Z/X \right)_{\astrosun} = 0.0178$, and the older mixture by \citep[][ GS98]{grevesse98} with a lower metallicity, $\left(Z/X \right)_{\astrosun} = 0.0229$, in order to study the impact of the solar chemical composition uncertainty in our results.


The Sun is evolved from the Zero Age Main Sequence (ZAMS) in a galactic halo with a constant dark matter energy density until its present age. For each series of WIMP parameters, the model is calibrated by automatic adjustment of the convection mixing length  parameter and Helium abundance, until it reaches a precision of $10^{-5}$ of the present luminosity, radius and mass. The calibration process takes an average of 10 iterated full-runs to obtain the desired precision. Dark matter capture, thermal annihilation (eq. \ref{eq:annInt}) and energy transport are computed at each time step, in order to consistently achieve the values expected today.\\  

\section{Results}
\label{sec:DMresults}

In this section we present the results obtained for thermally annihilating dark matter in the Sun for different WIMP masses, Spin-Independent (SI) and Spin-Dependent (SD) scattering cross-sections. We used a dark matter density of $\rho_{\chi} = 0.38 \ \text{GeV} \ \text{cm}^{-3}$~\citep{catena10}. We assume the standard value for the local orbital speed of the Sun $v_{\astrosun} = 220 \ \text{km} \ \text{s}^{-1}$~\citep{kerr86}, which results in a velocity dispersion for the standard Maxwellian dark matter halo of $\bar{v}\simeq \sqrt{3/2} v_{\astrosun} \simeq 270 \ \text{km} \ \text{s}^{-1}$.

\subsection{The Isothermal Limit}
\label{sec:resultsDist}

\begin{figure*}[!htb]
	\centering
	\subfloat[Spin dependent]{
		\includegraphics[width=\columnwidth]{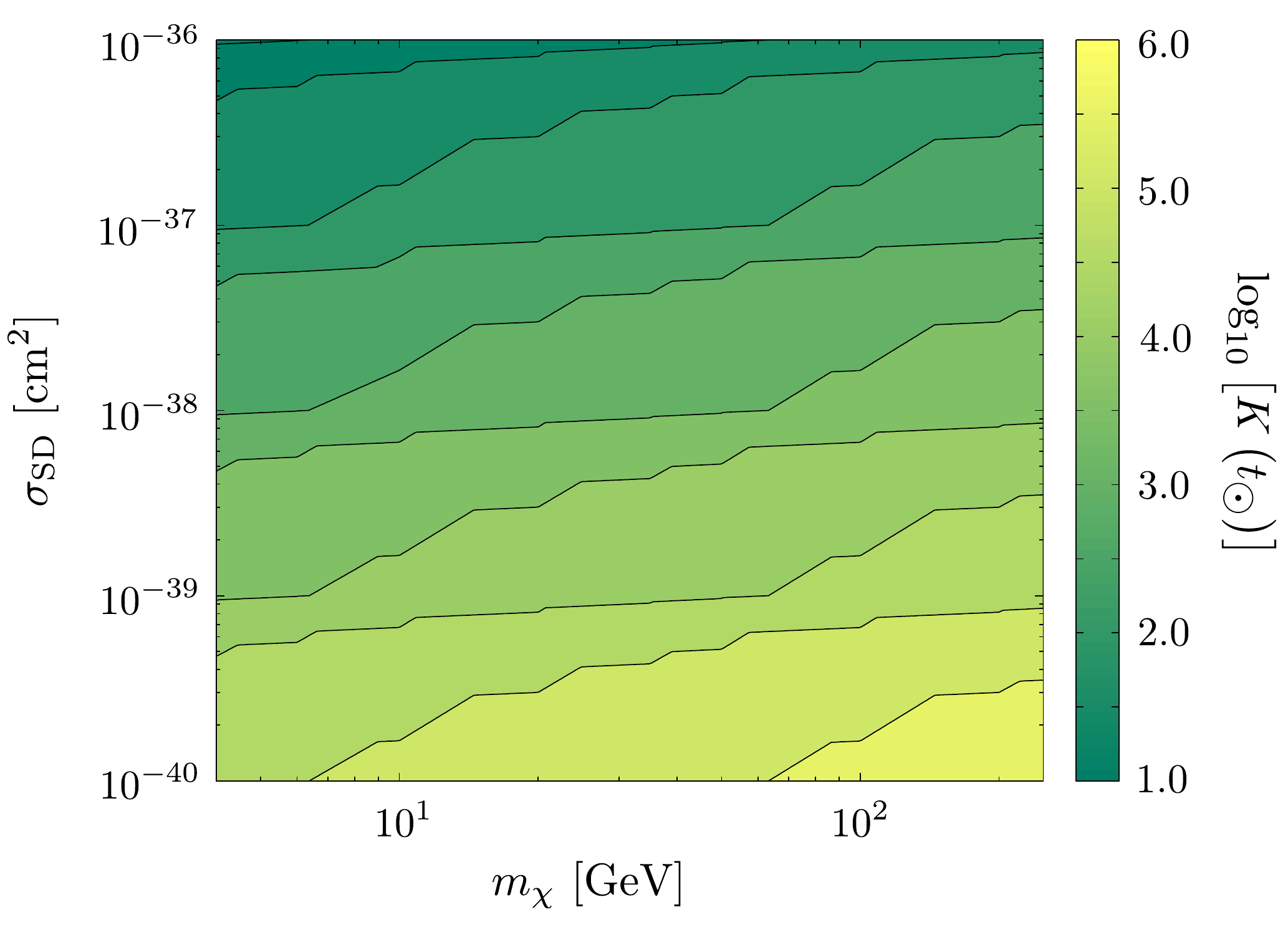}}
	\hfill
	\subfloat[Spin independent]{
		\includegraphics[width=\columnwidth]{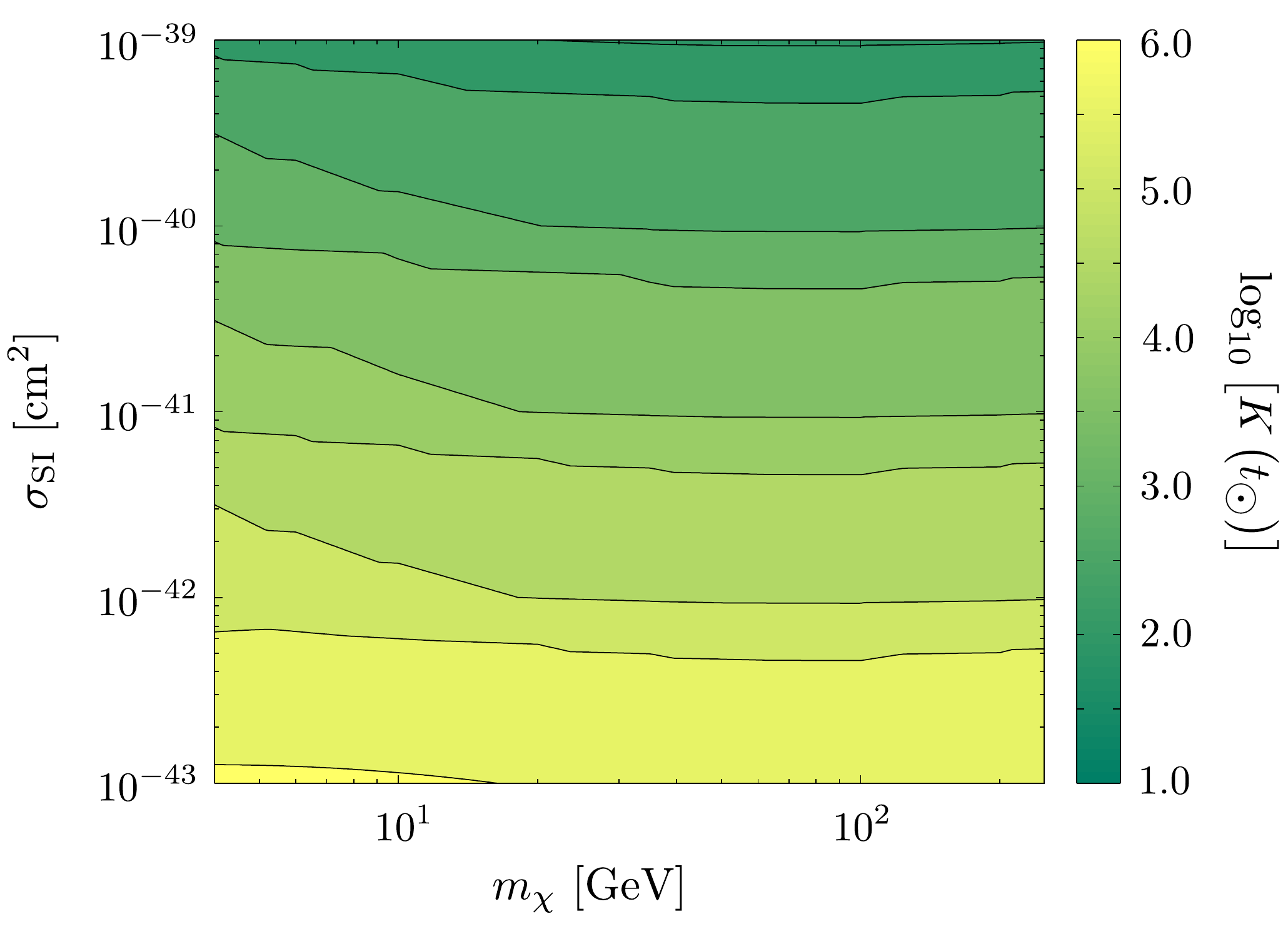}}
	\caption{The Knudsen number (\ref{eq:knudsen}) for different WIMP parameters. Colour scale is logarithmic.}
	\label{fig:knudsen}
\end{figure*}

To study how WIMPs are distributed in the solar medium, we computed the Knudsen Number (eq. \ref{eq:knudsen}) at the present age, $K(t_{\astrosun})$, in the WIMP parameter region of interest \ref{fig:knudsen}. 
Due to the low scattering cross-sections, WIMPs will orbit the Sun several times before scattering with nuclei, which means that in this region, $K$ is always larger than 1. Taking this into account, and the fact that WIMPs have thermalized in the Sun \citep{spergel85,griest87}, we can assume that the WIMP population in the Sun reflects a global temperature $T_{\chi}$, and which distribution is simply given by a Boltzmann distribution with state density defined by the gravitational potential $\phi(r)$,
\begin{equation}
n_{\chi}(r)\simeq n_{\chi, \text{iso}}(r) = N_{\chi} \frac{e^{-m_{\chi}\phi(r)/kT_{\chi}}}{\int_{\astrosun} e^{-m_{\chi}\phi(r')/kT_{\chi}} dr'^3}.
\end{equation}

Taking into account that WIMPs tend to strongly cluster on the Sun's core (for example, $r_{\chi}\simeq 0.05R_{\astrosun}$ for $m_{\chi}=4$ GeV and $r_{\chi}<0.01R_{\astrosun}$ for $m_{\chi}\gtrsim150$ GeV) it is accurate to approximate the potential $\phi(r)$ for $r<r_{\chi}$ by assuming a constant solar central density $\rho_{\text{c}}$, which yields $\phi(r)\propto \rho_{\text{c}}r^2$, resulting in a much simpler expression for the WIMP distribution,

\begin{equation}
n_{\chi}(r)\simeq N_{\chi} \frac{e^{-\frac{r^2}{r_{\chi}^2}}}{\pi^{\frac{3}{2}}r_{\chi}^3}.
\label{eq:DMdistSimp}
\end{equation}

In equation \ref{eq:DMdistSimp} we also used eq. \ref{eq:dmLenght} assuming that $T_\chi \simeq T_{\text{c}}(t)$, which is accurate for higher $m_{\chi}$. For lower masses, $m_{\chi}\approx 5 $ GeV, the error in this approximation is always lower than 7\% regarding the temperature, which is somewhat mitigated by the fact that WIMPs with lower masses will be close to equilibrium (see sec. \ref{sec:resultsEquilibrium}). Using equation \ref{eq:DMdistSimp}, we can solve the integral for annihilation coefficient (eq. \ref{eq:annInt}), which yields,
\begin{equation}
A_{\astrosun} = \frac{ T_{\text{c}}}{m_{\chi}} \frac{ b' }{4 \pi r_{\chi}^3}\left[ \sqrt{\frac{2}{\pi}} \text{Erf}\left( \sqrt{2}\frac{ R_{\astrosun}}{r_{\chi}}\right) - \frac{4}{\pi} \frac{R_{\astrosun}}{r_{\chi}} e^{-2\frac{R_{\astrosun}^2}{r_{\chi}^2}} \right]
\label{eq:annAna}
\end{equation}

where $R_{\astrosun}\simeq 6.96 \times 10^{10}$ cm, is the Sun radius. Taking into account that in our case $R_{\astrosun} \gg r_{\chi}$, we can simplify eq. \ref{eq:annAna} to the final annihilation coefficient expression,
\begin{equation}
A_{\astrosun} \simeq \frac{ T_{\text{c}}}{m_{\chi}} \frac{ b' }{\sqrt{8} \pi^{\frac{3}{2}} r_{\chi}^3}
\label{eq:annSimp}
\end{equation}
which is independent of the Sun radius, as expected.

\subsection{Annihilation Rate and Equilibrium Regime}
\label{sec:resultsEquilibrium}


Temperatures inside the Sun range from $T\simeq 10^7 \ - \ 10^4$ K, which means that the thermally averaged cross-section (eq. \ref{eq:annInt}) inside the Sun will be dramatically lower comparatively to the freeze-out epoch, that, as we saw in sec. \ref{sec:DMabundance}, happens for much higher temperatures. A lower annihilation cross-section, will allow further WIMP accretion, resulting in a larger number of WIMPs in the Sun relatively to the $s$-wave annihilation case. If the number of WIMPs in the Sun is in equilibrium, the total annihilation rate will be the same as the $s$-wave case, since in this regime, the annihilation rate depends exclusively on the capture rate (see eq. \ref{eq:annihilationRate}). However, a lower annihilation cross-section can cause the number of WIMPs to fall out of equilibrium, and in this case, the annihilation rate will be different from the case with standard $s$-wave annihilations.\\

\begin{figure*}[!htb]
	\centering
	\subfloat[Spin dependent]{
		\includegraphics[width=\columnwidth]{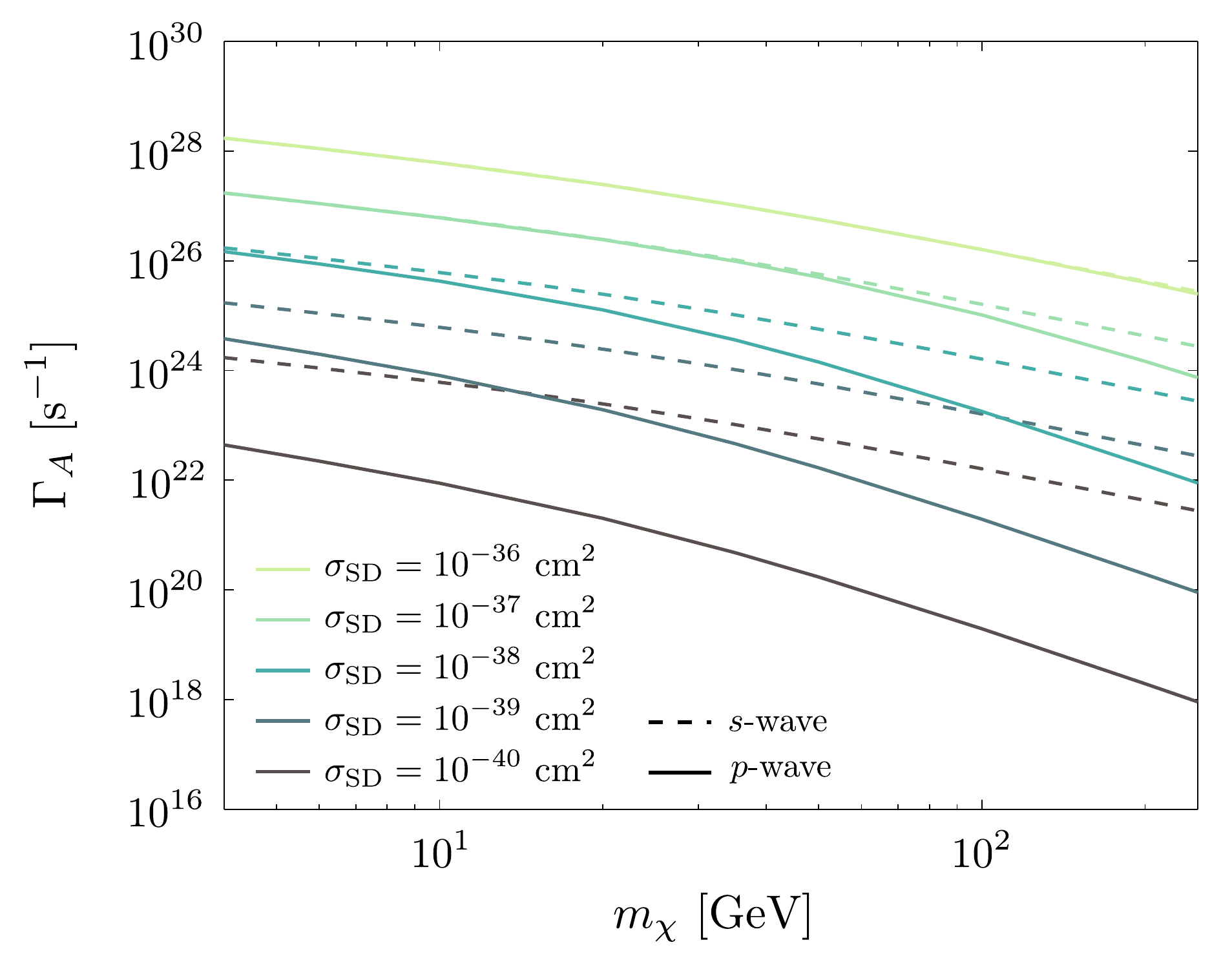}}
	\hfill
	\subfloat[Spin independent]{
		\includegraphics[width=\columnwidth]{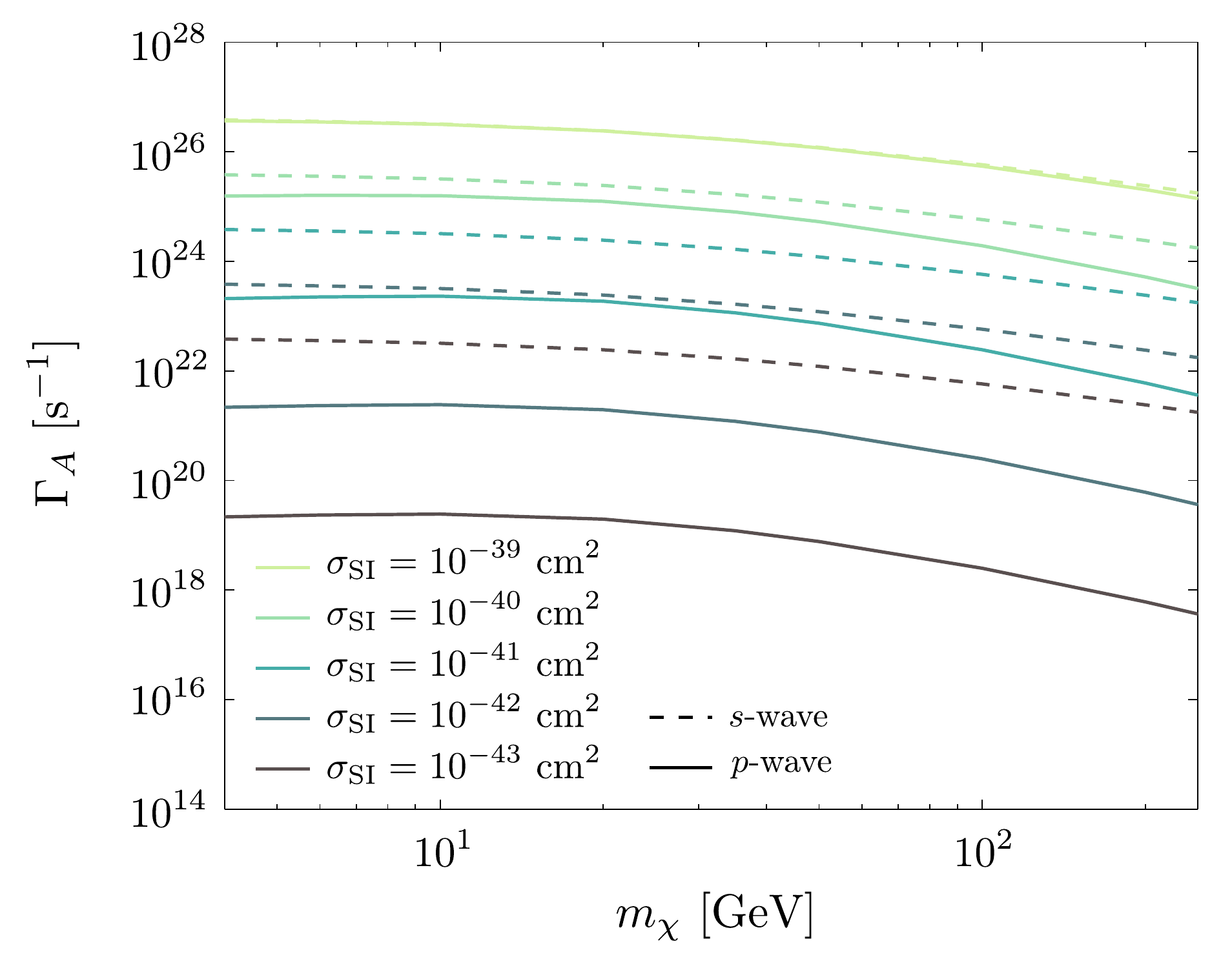}}
	\caption{Annihilation Rate for different SD and SI cross-sections. Dotted: Annihilation rate for models with $s$-wave annihilation. Full: Annihilation rate for models with $p$-wave annihilation.}
	\label{fig:annihilation}
\end{figure*}

\begin{figure*}[!htb]
	\centering
	\subfloat[Spin dependent]{
		\includegraphics[width=\columnwidth]{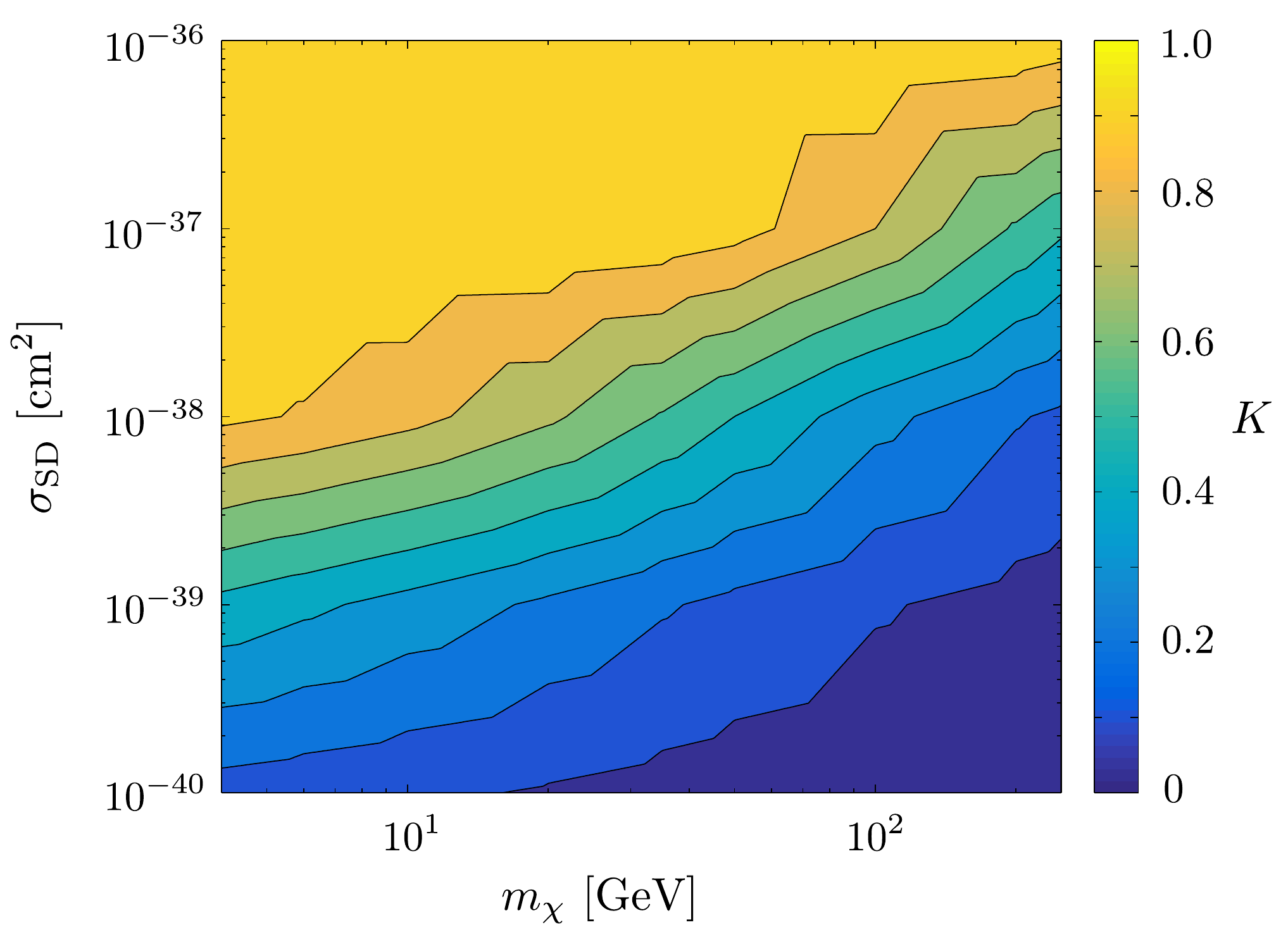}}
	\hfill
	\subfloat[Spin independent]{
		\includegraphics[width=\columnwidth]{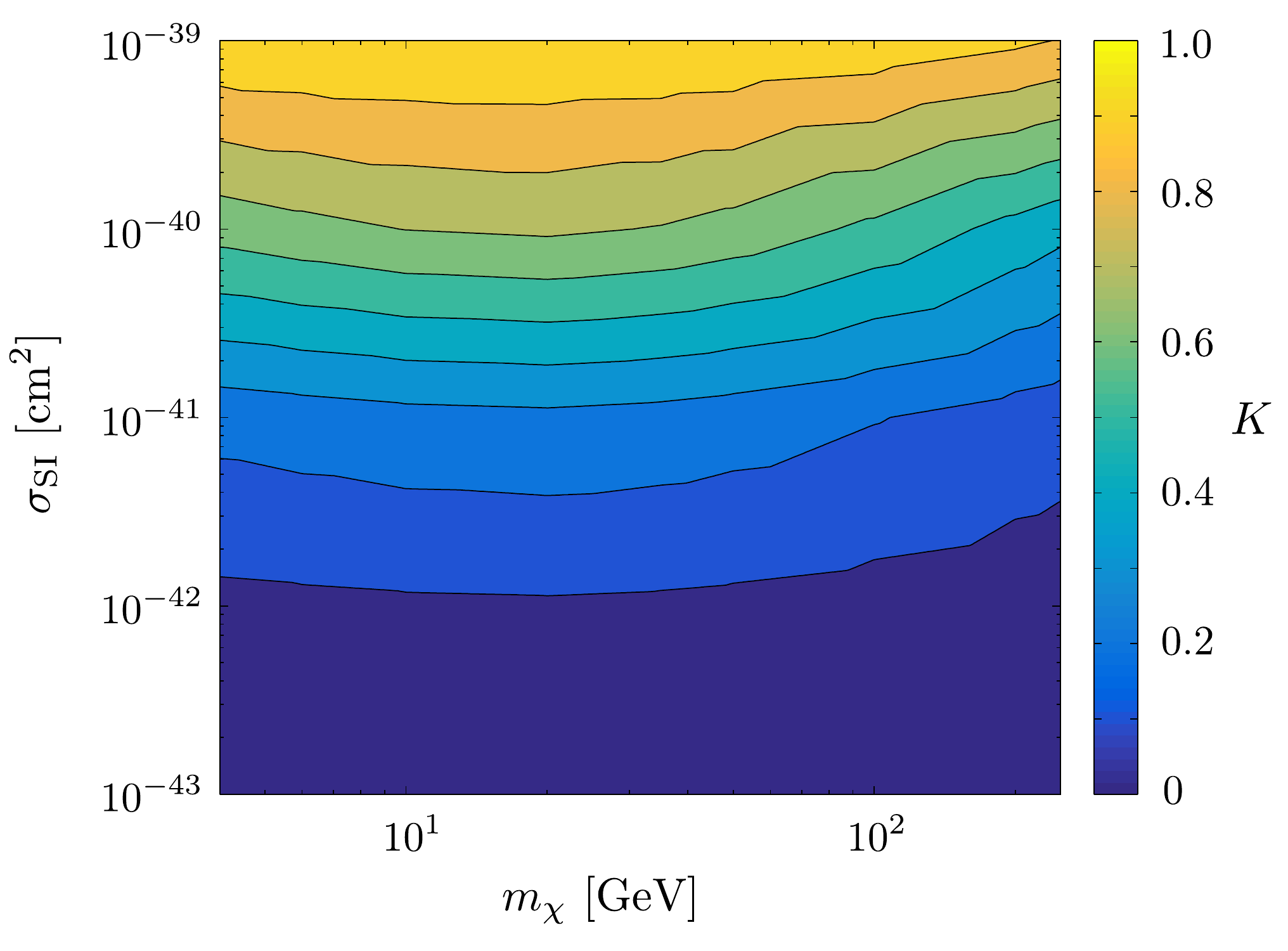}}
	\caption{Equilibrium parameter, $K$, for different Wimp parameters. Yellow zones are in equilibrium and the annihilation Rate is independent of the annihilation cross-section. Blue zones are not in equilibrium and the annihilation rate depends on the annihilation cross-section.}
	\label{fig:equilibrium}
\end{figure*}

In figure \ref{fig:annihilation} we computed the annihilation rate, for WIMPs isothermally distributed (eq. \ref{eq:DMdistSimp}) using the velocity dependent annihilation cross-section in eq. \ref{eq:annInt} for different WIMP masses and scattering cross-sections. For comparison, the annihilation rate for the $s$-wave case is also shown. As we can see, for larger scattering cross sections ($\sigma_{SD} = 10^{-36} \ \text{cm}^2$ and $\sigma_{SI} = 10^{-39} \ \text{cm}^2$) there is almost no difference between the two cases, since the number of WIMPs has achieved equilibrium in the two cases. However, as we increase the mass and decrease the scattering cross-section, there is a large difference between the $s$-wave and $p$-wave case. This is also visible in figure \ref{fig:equilibrium}, where we plotted the hyperbolic tangent in eq. \ref{eq:wimpNumber}, 
\begin{equation}
K \equiv \text{tanh}\left(t_{\astrosun}/ t_{\text{eq}}\right),
\end{equation}
which measures the departure from WIMP number equilibrium. Regions where the number of WIMPs is far from equilibrium, i.e. where $K\ll1$, the annihilation rate (fig. \ref{fig:annihilation}) will be lower in the $p$-wave case. In fact, for smaller scattering cross-sections ($\sigma_{SD} = 10^{-40} \ \text{cm}^2$ and $\sigma_{SI} = 10^{-43} \ \text{cm}^2$) and higher masses, the difference between the two annihilation cases can be of several orders of magnitude, which will result in more loose constraints, as we will show in the next section. \\





\subsection{New limits}
\label{sec:limits}
	
	WIMPs in the Sun will annihilate to SM particles, from which only neutrinos will be able to escape the Sun without loosing all of their energy, due to their weakly interacting nature. The resulting neutrino flux from WIMP annihilation in the Sun is given by
	\begin{equation}
	\Phi_{\nu} = \frac{\Gamma_A}{4 \pi r^2} \sum_{i} \text{BR}_i \int \frac{\text{d}N_{\nu}}{\text{d}E_{\nu}}\text{d}E_{\nu}
	\end{equation}
	where the sum is done over the $i$ possible annihilation channels, with Branching ratio $\text{BR}_i$ and spectra $\text{d}N_{\nu}/\text{d}E_{\nu}$. Note that the branching ratios for each annihilation channel are not known for a general WIMP model. However, most WIMPs annihilate predominantly to one particular state, such as $b \bar{b}$, $\tau^+ \tau^-$ and $W^+ W^-$. The neutrino signal can be detected on Earth using large Cerenkov detectors, such as the \textsc{IceCube}, in the South Pole, and the \textsc{Super-Kamiokande}, in the \textit{Kamioka Mine}, Japan.
	To compute the neutrino flux measured at the Earth, we used \textsc{WimpSim} ~\citep{blennow08} (integrated with our stellar evolution code), which uses an event-based framework. \textsc{WimpSim} computes WIMP annihilation to SM particles, and their subsequent hadronization or decay to neutrinos using \textsc{Pythia 6.400} ~\citep{pythia06}. After that, it propagates the neutrino signal through the solar medium and vacuum, using a three-flavour oscillation framework and taking into account neutrino charged currents (CC) and neutral currents (NC) with the solar nuclei, which will have impact in the neutrino spectra for higher energies. \\
	
	\begin{table*}[t]
		\centering
		\scriptsize
		\caption{Upper limits on the SD and SI cross-sections from the Super-Kamiokande detector for $s$ and $p$-wave annihilation. The 90 \% upper limit on the muon-neutrino flux, $\Phi_{\nu_{\mu}} $, was taken from ref.~\citet{SK15}. The annihilation rate $\Gamma_A$ for each case is also shown for comparison. }
		\begin{tabular}{llcccclcc}
			&  & \multicolumn{1}{l}{} &  & \multicolumn{2}{c}{\normalsize{\textsc{$s$-wave}}} &  & \multicolumn{2}{c}{\normalsize{\textsc{$p$-wave}}} \\ \cline{5-6} \cline{8-9} 
			\multicolumn{1}{c}{$m_{\chi}$} & \multicolumn{1}{c}{\multirow{2}{*}{Channel}} & $\Phi_{\nu_{\mu}} $ & $\Gamma_A$ & $\sigma_{\text{SD}}$ & $\sigma_{\text{SI}}$ &  & $\sigma_{\text{SD}} $ & $\sigma_{\text{SI}}$ \\
			\multicolumn{1}{c}{(GeV)} & \multicolumn{1}{c}{} & $(10^{12} \ \mathbf{\text{\textbf{km}}^{-2}} \ \text{y}^{-1})$ & $(\text{s}^{-1})$ & $ (10^{-40} \ \mathbf{\text{\textbf{cm}}^2})$ & $(10^{-42} \ \mathbf{\text{\textbf{cm}}^2})$ &  & $(10^{-40} \ \mathbf{\text{\textbf{cm}}^2})$ & $(10^{-41} \ \mathbf{\text{\textbf{cm}}^2})$ \\ \hline \hline
			4 & $\tau^- \tau^+$ & 150 & $2.85 \times 10^{24}$ & 1.65 & 7.50 &  & 7.71 & 2.55 \\ \hline
			\multirow{2}{*}{6} & $b \bar{b}$ & 294 & $1.42 \times 10^{25}$ & 12.8 & 39.8 &  & 22.8 & 8.98 \\
			& $\tau^- \tau^+$ & 70.8 & $1.36 \times 10^{24}$ & 1.23 & 3.82 &  & 7.14 & 1.65 \\ \hline
			\multirow{2}{*}{10} & $b \bar{b}$ & 140 & $6.78 \times 10^{24}$ & 11.1 & 21.2 &  & 22.8 & 4.80 \\
			& $\tau^- \tau^+$ & 31.0 & $6.05 \times 10^{23}$ & 0.99 & 1.89 &  & 7.70 & 1.22 \\ \hline
			\multirow{2}{*}{20} & $b \bar{b}$ & 53.1 & $2.65 \times 10^{24}$ & 10.8 & 10.9 &  & 27.6 & 2.81 \\
			& $\tau^- \tau^+$ & 13.2 & $2.62 \times 10^{23}$ & 1.07 & 1.08 &  & 10.5 & 1.05 \\ \hline
			\multirow{2}{*}{50} & $b \bar{b}$ & 19.8 & $9.99 \times 10^{23}$ & 17.6 & 8.30 &  & 73.1 & 2.59 \\
			& $\tau^- \tau^+$ & 2.67 & $5.33 \times 10^{22}$ & 0.94 & 0.44 &  & 12.3 & 0.96 \\ \hline
			\multirow{2}{*}{100} & $b \bar{b}$ & 7.54 & $3.77 \times 10^{23}$ & 23.4 & 6.53 &  & 118 & 2.67 \\
			& $\tau^- \tau^+$ & 0.70 & $1.44 \times 10^{22}$ & 0.89 & 0.25 &  & 16.3 & 0.95 \\ \hline
			\multirow{2}{*}{200} & $b \bar{b}$ & 2.81 & $1.45 \times 10^{23}$ & 34.0 & 6.05 &  & 178 & 3.44 \\
			& $\tau^- \tau^+$ & 0.19 & $4.12 \times 10^{21}$ & 0.97 & 0.17 &  & 29.0 & 0.97\\ \hline \hline
		\end{tabular}
		\label{tab:sk}
	\end{table*}
	
	\begin{table*}[t]
		\centering
		\scriptsize
		\caption{Upper limits on the SD and SI cross-sections from the IceCube detector for $s$ and $p$-wave annihilation. The 90 \% upper limit on the muon flux, $\Phi_{\mu} $, was taken from ~\citet{IC13}. The annihilation rate $\Gamma_A$ for each case is also shown for comparison. }
		\begin{tabular}{llcccclcc}
			&  & \multicolumn{1}{l}{} &  & \multicolumn{2}{c}{\normalsize{\textsc{$s$-wave}}} &  & \multicolumn{2}{c}{\normalsize{\textsc{$p$-wave}}} \\ \cline{5-6} \cline{8-9} 
			\multicolumn{1}{c}{$m_{\chi}$} & \multicolumn{1}{c}{\multirow{2}{*}{Channel}} & $\Phi_{\mu} $ & $\Gamma_A$ & $\sigma_{\text{SD}}$ & $\sigma_{\text{SI}}$ &  & $\sigma_{\text{SD}} $ & $\sigma_{\text{SI}}$ \\
			\multicolumn{1}{c}{(GeV)} & \multicolumn{1}{c}{} & $(10^{4} \ \mathbf{\text{\textbf{km}}^{-2}} \ \text{y}^{-1})$ & $(\text{s}^{-1})$ & $ (10^{-39} \ \mathbf{\text{\textbf{cm}}^2})$ & $(10^{-41} \ \mathbf{\text{\textbf{cm}}^2})$ &  & $(10^{-39} \ \mathbf{\text{\textbf{cm}}^2})$ & $(10^{-40} \ \mathbf{\text{\textbf{cm}}^2})$ \\ \hline \hline
			20 & $\tau^- \tau^+$ & 9.27 & $2.38 \times 10^{25}$ & 9.69 & 9.83 &  & 14.3 & 1.45 \\ \hline
			\multirow{2}{*}{35} & $b \bar{b}$ & 10.4 & $1.70 \times 10^{26}$ & 164 &  103  &  & 169 &  10.5   \\
			& $\tau^- \tau^+$ & 1.21 & $9.61 \times 10^{23}$ & 0.93 & 0.58 &  & 3.30 & 0.20 \\ \hline
			\multirow{2}{*}{50} & $b \bar{b}$ & 1.80 & $1.68 \times 10^{25}$ & 29.6 & 14.0 &  & 38.4 & 1.92 \\
			& $\tau^- \tau^+$ & 0.28 & $1.17 \times 10^{23}$ & 0.21 & 0.10 &  & 1.64 & 0.11 \\ \hline
			\multirow{2}{*}{100} & $W^+W^-$ & 0.12 & $3.24 \times 10^{22}$ & 0.20 & 0.06 &  & 2.54 & 0.10 \\
			& $b \bar{b}$ & 0.59 & $1.78 \times 10^{24}$ & 11.1 & 3.09 &  & 24.2 & 0.93 \\ \hline
			\multirow{2}{*}{250} & $W^+W^-$ & 0.04 & $2.61 \times 10^{21}$ &  0.09   & 0.01 &  & 3.54 & 0.10 \\
			& $b \bar{b}$ & 0.15 & $1.23 \times 10^{23}$ & 4.16 & 0.70 &  & 24.0 & 0.44 \\ \hline \hline
		\end{tabular}
		\label{tab:ic}
	\end{table*}
	
		\begin{figure*}[!htb]
			\centering
			\subfloat[Spin dependent]{
				\includegraphics[width=\columnwidth]{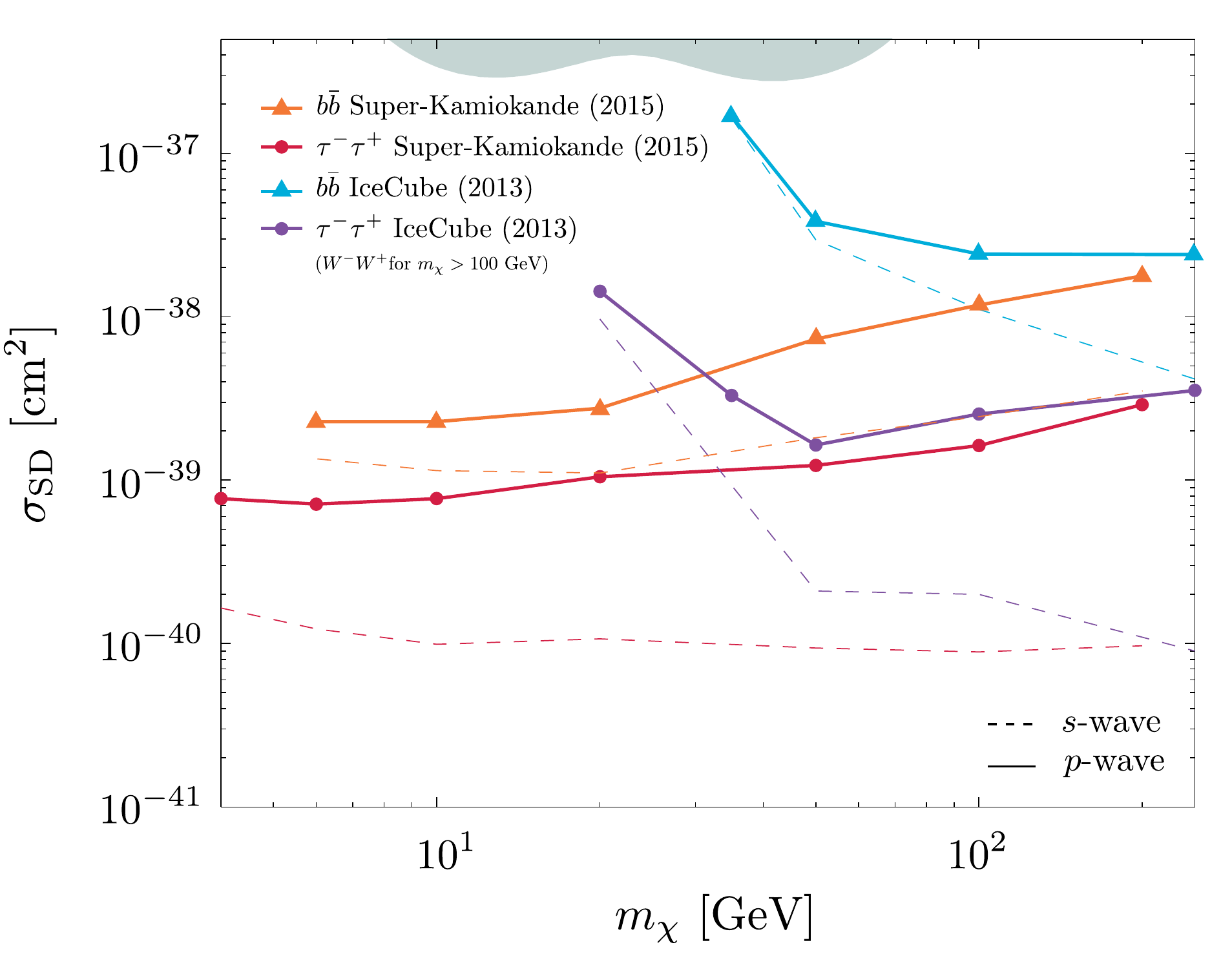}}
			\hfill
			\subfloat[Spin independent]{
				\includegraphics[width=\columnwidth]{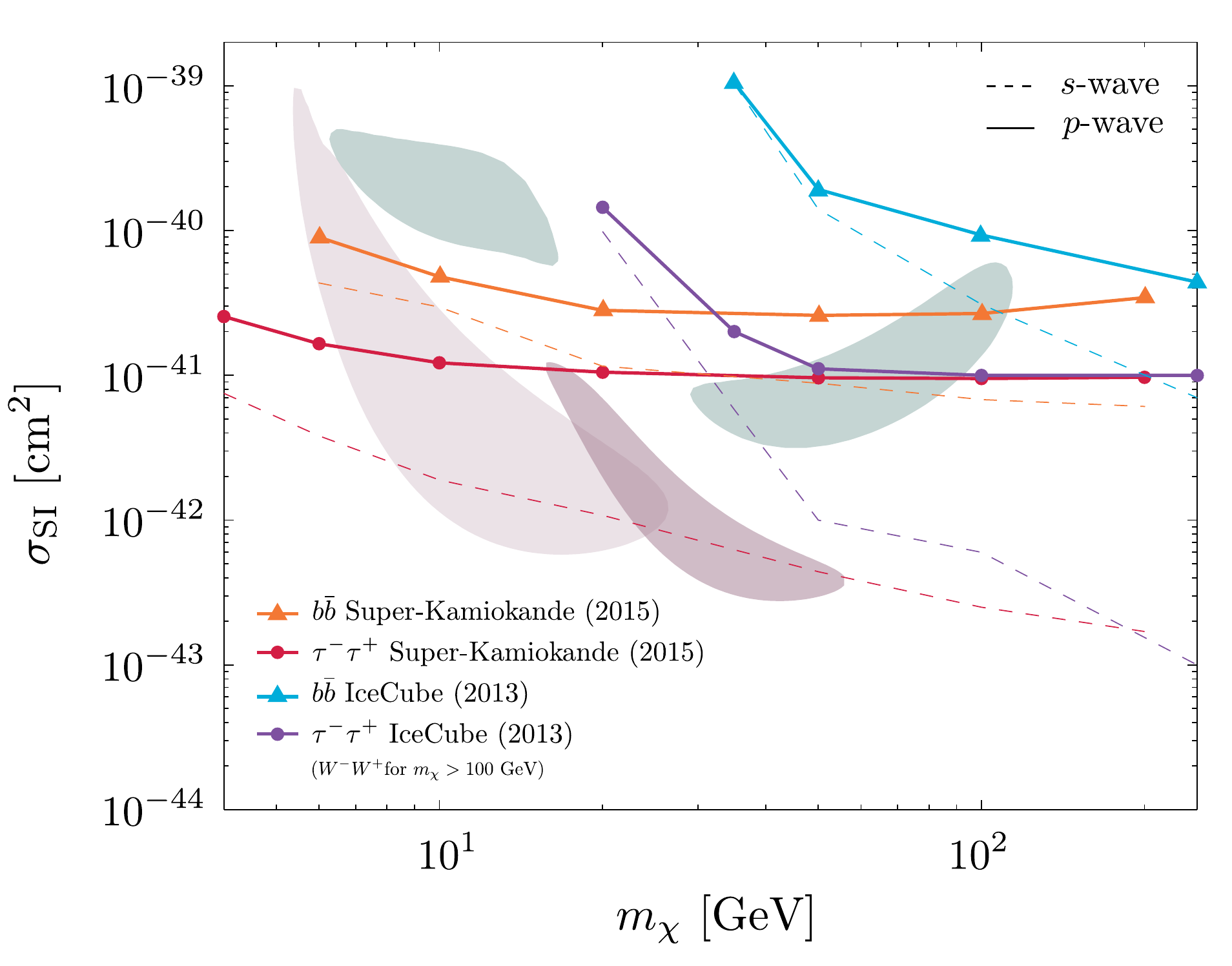}}
			\caption{Limits in the SD and SI scattering cross-section placed by the Super-Kamiokande and IceCube neutrino detectors. Dashed: Limits for the $s$-wave annihilation case (standard). Solid: Limits for $p$-wave annihilation. The favoured regions from different direct detection are also shown. Pink: \textsc{CDMS II} Si at 95 \% C.L. ~\citep{agnese13}; Green: \textsc{DAMA/LIBRA} at $3\sigma$ C.L. ~\citep{bernabei08}; Purple: \textsc{CRESSTII} at $2\sigma$ C.L. ~\citep{angloher12}.}
			\label{fig:constraints}
		\end{figure*}
		
	The upper-limits on the neutrino flux from WIMP annihilation in the \textsc{IceCube} and \textsc{Super-Kamiokande} were converted to annihilation rates using \textsc{WimpSim} (see tables \ref{tab:sk} and \ref{tab:ic}). The annihilation rates, $\Gamma_A$, where converted into upper limits on the scattering cross-section (SD and SI) using our stellar evolution code to compute Dark Matter capture and $p$-wave annihilation. We also computed the upper limits on the scattering cross-section for WIMPs annihilation through $s$-wave for comparison reasons. In fig. \ref{fig:constraints} we plotted the upper limits for $s$ and $p$-wave annihilation, as well as limits from different direct detection experiments (see figure caption for references). \\
	
	As expected, for regions where $t_{eq}\ll t_{\astrosun}$, the limits on the scattering cross-section for WIMP models with dominant $p$-wave annihilation are coincident with the limits for the standard $s$-wave case. However, as we increase $m_{\chi}$ and decrease the scattering cross-section, the number of WIMPs will fall out of equilibrium resulting in smaller annihilation rates, as shown in sec. \ref{sec:resultsEquilibrium}, which will convert to upper limits on the SD and SI scattering cross-sections above the standard case, specially for higher masses.\\
	
	\section{Discussion}
	
	In sec.~\ref{sec:limits} we obtained the limits on the scattering cross-section for p-wave annihilating WIMPs, placed by the Super-Kamiokande and IceCube neutrino telescopes. We also computed the limits for the standard case of s-wave annihilation, which are in fair agreement with the results obtained by previous analysis, including the experimental collaborations' ~\citep{IC13,SK15}. It should be noted that our constraints for the s-wave case are slightly below than those obtained by the experimental collaborations, which is simply due to the fact that we used a newly determined Dark Matter local halo density of $ \rho_\chi = 0.38 \ \text{GeV cm}^{-3}$\citep{catena10}, while in the other analysis the traditional density $ \rho_\chi = 0.30 \ \text{GeV cm}^{-3}$ is used. Despite being presented in a separate fashion, the constraints presented in sec. \ref{sec:limits} can be interpreted as the limiting cases of WIMPs annihilating with both s and p-wave contributions.
	
	A previous analysis by~\citet{kappl11} also included the computation of the limits on the scattering cross-section for light WIMPs ($m_\chi \lesssim 20$ GeV for SI and $m_\chi \lesssim 80$ GeV for SD) with pure p-wave annihilation. Albeit not including the most recent Super-Kamiokande run (SK IV, 2008-2012), the analysis by~\citet{kappl11} already showed that the p-wave limits are considerably less stringent than those from pure s-wave annihilations, which stems from the fact that the WIMP annihilation cross-sections within the Sun are smaller than those during the freeze-out epoch. In this work, we extended and improved this analysis by employing the most recent limits to this date by the Super-Kamiokande and IceCube experiments to a wider range of WIMP masses, which are currently responsible for the strongest bounds on the spin-dependent cross-section for WIMPs with dominant s-wave annihilation. Similarly to \cite{kappl11}, in the SI case, we found that the favoured regions by the CDMS ~\citep{agnese13}, DAMA/LIBRA ~\citep{bernabei08} and CRESSTII ~\citep{angloher12} remain viable for WIMPs annihilating mainly through the $\tau^+ \tau^-$ or $b \bar{b}$ annihilation channel with pure dominant p-wave annihilation. On the other hand, for the SD scenario, the DAMA/LIBRA favoured region is ruled out for both s and p-wave annihilations.
	
	The result obtained also shows that as experiments are built to further constrain the neutrino signal from WIMP annihilation, it will be continuously harder to survey lower scattering cross-sections for models with pure $p$-wave annihilation, since we will enter the parameter region where the number of WIMPs is far from equilibrium (see fig. \ref{fig:equilibrium}). In this limit, where $t_{eq} \gg t_{\astrosun}$, the number of WIMPs in the Sun can be approximated by
	
	\begin{equation}
	N_{\chi}(t_{\astrosun})= \sqrt{\frac{C_{\astrosun}}{A_{\astrosun}}} \ \text{tanh}(\sqrt{C_{\astrosun}A_{\astrosun}} t_{\astrosun}) \simeq C_{\astrosun}t_{\astrosun},
	\end{equation}
	
	resulting in 
	
	\begin{equation}
	\Gamma_A \simeq \frac{1}{2}\left(C_{\astrosun}t_{\astrosun}\right)^2 A_{\astrosun}, \qquad \left( C_{\astrosun} \propto \sigma_{\text{SD/SI}} \right),
	\end{equation}
	
	which means that if we are able to lower the upper limit in the Neutrino flux by a factor of 10, it will result in lowering the upper limit on the scattering cross-section by a factor of approximately 3, for models with pure $p$-wave annihilation.\\
	
	The limits on the scattering cross-section presented in~\ref{sec:DMresults} are susceptible to various uncertainties, mainly stemming from the current uncertainty on the capture rate of WIMPs from the Sun, which in turn is sensible to the uncertainty on the local dark matter density $\rho_\chi$ and the velocity distribution of WIMPs in the Milky Way. Regarding the former, independent determinations have shown that the local dark matter can be at least a factor of two larger than the value used here~\citep{salucci10,pato10,garbari11,garbari12}. This uncertainty can be simply accounted, since the capture rate is directly proportional to $\rho_{\chi}$. The same however is not true for the uncertainty associated with the velocity distribution of WIMPs in the Milky way, which contributes non-trivially for the capture rate. Motivated by the growing tensions between direct detection results, recent N-body simulations have shown that the assumption of a Maxwellian velocity distribution can overestimate the number of WIMPs in the high velocity tail of the velocity distribution \citep{mao13,mao14}, on which the direct detection rates depend strongly. It is important to note these results stem from dark matter only simulations, and more recent distributions inferred from simulations including baryons somewhat reduce the discrepancy with the standard Maxwellian distribution \citep{kelso16,sloane16}. Moreover, the capture rate is only sensible to the low velocity tail of the WIMP velocity distribution, given that low velocity WIMPs are more prone to be gravitationally captured by the Sun. \citet{choi14}~have studied the impact of the WIMP velocity distribution on the capture rate, and they found that for light WIMPs ($m_{\chi}\simeq 20$ GeV) there is an uncertainty on the capture rate of 24\% (17\%) for SD (SI) interactions, while for heavier WIMPs ($m_{\chi}\simeq 500$ GeV), the impact is more dramatic, with uncertainties up to 50 \% . These results include both the uncertainty on the WIMP velocity distribution and on the parameters $v_{\astrosun}$ and $\bar{v}$. The error induced by the capture rate uncertainty on the limits from neutrino telescopes will be different for the s and p-wave cases. For the s-wave case, since $\Gamma_A \propto C_{\astrosun}$, there is a direct correspondence between uncertainties \cite[see][for a review on the uncertainties of the s-wave limits from neutrino telescopes]{danninger14}. For the p-wave scenario however, as shown in sec.~\ref{sec:resultsEquilibrium}, the number of WIMPs will generally not be in equilibrium. In the extreme case $\Gamma_A \propto C_{\astrosun}^2$, and thus we can assume as a conservative estimation for the error of the scattering cross-section limits in the p-wave case as $\delta \sigma_{\text{SI/SD}} \simeq 2 \delta C_{\astrosun, \text{SI/SD}}$
	
 We also studied how the current uncertainty on the overall solar models affects our results, by repeating the computation of the p-wave limits from neutrino telescopes for meticulously chosen benchmark values of $m_\chi$, $\sigma_{\text{SD}}$ and $\sigma_{\text{SI}}$ using the older mixture of abundances by~\citet{grevesse98} (GS98). Again, we found that the largest source of uncertainty is the capture rate $C_{\astrosun}$, which for SD (only scattering with hydrogen is taken into account) scattering is larger for AGSS09 ($\sim 4 \%$ independent of $m_{\chi}$), and for SI (scattering with all the elements) scattering is larger ( $\sim 14 \%$ for $m_{\chi}=10$ GeV and $\sim 19 \%$ for $m_{\chi}=200$ GeV) for GS98. This difference, which is a direct consequence of the difference in $\left(Z/X \right)_{\astrosun}$ between models (see the discussion in~\ref{sec:DMmodel}), can have impact in the upper limits for $\sigma_{\text{SD/SI}}$ since the annihilation rate $\Gamma_A$ depends on the Number of Wimps $N_{\chi}$, which in turn depends on $C_{\astrosun}$. The difference in the central temperature between solar models evolved with the mixture by GS98 and AGSS09, ($\sim 2 \%$ higher $T_{\text{c}}$ for GS98) will produce an uncertainty in the annihilation coefficient (eq.~\ref{eq:annSimp}) which is sub-dominant compared to the uncertainty in the capture Rate. In the overall picture, for solar models with GS98, upper limits on the SD cross-section will be $\sim 4 \%$ more relaxed, while upper limits in the SI cross-section will be tighter ($\sim 5\%$ for annihilation to $\tau^- \tau^+$ and $\sim 20 \%$ for $b\bar{b}$).
	
	\section{Conclusions}
	
	Dark Matter particles trapped in the Sun will annihilate and create a neutrino signal that can be used to survey its properties. In this article we studied the neutrino emission for a simple model in which the main contribution for the annihilation comes from the $p$-wave channel, i.e. a velocity dependent cross-section. To obtain the annihilation cross-section we numerically solved the Boltzmann equation for the density of a relic particle satisfying the Dark Matter abundance as measured today from the CMB. To convert the upper limits on the neutrino flux from the \textsc{IceCube} and \textsc{SuperKamiokande} detectors to upper limits on the WIMP scattering cross-section we used a robust stellar evolution code to model the Sun including Dark Matter capture and annihilation. Assuming that WIMPs distribute isothermally in the Sun's core, we derived an analytical expression for the annihilation coefficient which is directly proportional to the solar central temperature.\\
	
	Differently from the usual case adopted in literature, where WIMPs annihilate through a velocity independent constant annihilation cross-section ($s$-wave channel), the neutrino signal will be directionally proportional to the annihilation coefficient, resulting in upper limits on the scattering cross-section of at least one order of magnitude above the $s$-wave case, which reduces the tension with results from other detection experiments.\\
	
	We also studied the impact of the current uncertainty on solar models (mainly due to the imposition of different solar heavy elements mixtures) in our results. We found out that models with a higher metallicity to hydrogen ratio have an enhancement of $\sim 20 \%$ on the Capture rate for Spin Independent scattering, while models with lower $\left(Z/X \right)_{\astrosun}$ capture $\sim 5\%$ more Dark Matter for Spin Dependent scattering. This variations can reflect a maximum of $\sim 20 \%$ uncertainty on the upper limits for the scattering cross-section for both $s$ and $p$-wave annihilations.\\

\acknowledgments
We would like to acknowledge the authors of the numerical packages used in this work, namely Morel and Lebreton for the stellar evolution code, \textsc{CESAM}; Eug\'enio and Casanellas for including all the Dark Matter framework in the code; and finally Blennow, Edsj\"o and Ohlsson for \textsc{WimpSim}. We would also like to thanks the anonymous referee for the insightful comments that helped improve this article.

\bibliographystyle{aasjournal} 

\begin{thebibliography}{}
\bibitem[Aartsen et al. (2013)]{IC13} Aartsen, M.G. et al. (\textsc{IceCube} Collaboration) 2013, \prl, 110, 131302
\bibitem[Ade et al. (2015)]{planck15} Ade, P.A.R et. al (\textsc{Planck} Collaboration) 2015, e-Print: arXiv:1502.01589.
\bibitem[Agnese et al. (2013)]{agnese13} Agnese, R. et al. (\textsc{CDMS} Collaboration) 2013, \prl, 111, 251301
\bibitem[Akerib et al. (2014)]{akerib15} Akerib, D.S. et al. (\textsc{LUX} Collaboration) 2014, \prl, 112, 091303
\bibitem[Angloher et al. (2012)]{angloher12} Angloher, G., Bauer, M., Bavykina, I., Bento, A., Bucci, C. et al. 2012, Eur. Phys. J., C72, 1971
\bibitem[Aprile et al. (2012)]{aprile12} Aprile, E. et al. (\textsc{XENON100} Collaboration) 2012, \prl, 109, 181301
\bibitem[Asplund et al. (2009)]{asplund09} Asplund, M., Grevesse, N., Sauval, A.J. \& Scott, P. 2009, Annual Review of Astronomy and Astrophysics
\bibitem[Bahcall \& Pinsonneault (2004)]{bahcall04} Bahcall, J.N. \& Pinsonneault, M.H. 2004, \prl, 92, 121301
\bibitem[Bahcall et al. (2005)]{bahcall05} Bahcall, J.N., Serenelli, A.M. \& Basu, S. 2005, \apj, 621, L85
\bibitem[Bahcall et al. (2006)]{bahcall06} Bahcall, J.N., Serenelli, A.M. \& Basu, S. 2006, \apjs, 165, 400
\bibitem[Basu \& Antia (2004)]{basu04}Basu, S. \& Antia, H.M. 2004, \apj, 606, L85
\bibitem[Bernabei et al. (2008)]{bernabei08} Bernabei, R. et al. (\textsc{DAMA/LIBRA} Collaboration) 2008, Eur. Phys. J., C56, 333
\bibitem[Bertone et al. (2005)]{bertone05} Bertone, G., Hooper, D. \& Silk, J. 2005, Physics Reports, 405, 279
\bibitem[Busoni et al. (2013)]{busoni13} Busoni, G., De Simone, A., Huang, W. 2013, JCAP, 07, 010
\bibitem[Blennow et al. (2008)]{blennow08} Blennow, M., Edsj\"o, J. \& Ohlsson, T. 2008, JCAP, 01, 021
\bibitem[Caffau et al. (2011)]{caffau11}Caffau, E., Ludwig, H.G., Steffen, M., Freytag, B. \& Bonifacio, P. 2011, Sol. Phys., 268, 255
\bibitem[Catena \& Ullio (2010)]{catena10} Catena, R. \& Ullio, P. 2010, JCAP, 08, 004
\bibitem[Choi et al. (2014)]{choi14} Choi, K., Rott, C. \& Itow, Y. 2014, JCAP, 05, 049
\bibitem[Choi et al. (2015)]{SK15} Choi, K. et. al (Super-Kamiokande Collaboration) 2015, \prl, 114, 141301
\bibitem[Danninger \& Rott (2014)]{danninger14} Danninger, M. \& Rott, C. 2014, Physics of the Dark Universe, 5, 35
\bibitem[Dent et al. (2010)]{dent10} Dent, J.B., Dutta, S., Scherrer, R.J. 2010, Physics Letters B, 687, 275 
\bibitem[Drees \& Nojiri (1992)]{drees92} Drees, M.,
    \& Nojiri, M. M.  1992, 1992nrdm.book.....D
\bibitem[Gaisser et al. (1986)]{gaisser86} Gaisser, T.K., Steigman, G., Tilav, S. 1986, \prd, 34, 2206
\bibitem[Garbari et al. (2011)]{garbari11} Garbari, S., Read, J.I. \& Lake, G. 2011, Mon. Not. R. Astron. Soc., 416, 2318
\bibitem[Garbari et al. (2012)]{garbari12} Garbari, S., Liu, C., Read, J.I. \& Lake, G. 2012, Mon. Not. R. Astron. Soc., 425, 1445
\bibitem[Goldberg (1983)]{goldberg83} Goldberg, H. 1983, \prl, 50, 1419
\bibitem[Gould \& Raffelt (1990)]{gould90} Gould, A. \& Raffelt, G. 1990, \apj, 352, 654
\bibitem[Grevesse et al. (1998)]{grevesse98} Grevesse, N. \& Sauval, A.J. 1998, Space Sci. Rev., 85, 161
\bibitem[Griest \& Seckel (1987)]{griest87} Griest, K., Seckel, D. 1987, Nuclear Physics B, 283, 681
\bibitem[Haxton et al. (2013)]{haxton13} Haxton, W.C., Hamish Robertson, R.G. \& Serenelli, A.M. 2013, Annu. Rev. Astron. Astrophys., 51, 21
\bibitem[Jungman et al. (1996)]{jungman96} Jungman, G., Kamionkowski, M., \& Griest, K.  1996, Physics Reports,
267, 195
\bibitem[Kappl et al. (2011)]{kappl11} Kappl, R. \& Winkler, M.W. 2011, Nucl. Phys. B, 850, 505
\bibitem[Kelso et al. (2016)]{kelso16} Kelso, C., Savage, C., Valluri, M., Freese, K., Stinson, G.S. \& Bailin,J. 2016, e-Print: arXiv:1601.04725
\bibitem[Kerr \& Lynden-Bell (1986)]{kerr86} Kerr, F.J. \& Lynden-Bell, D. 1986, Mon. Not. R. Astron. Soc., 221, 1023
\bibitem[Kim and Lee (2007)]{kim07} Kim, H.G., Lee, K.Y. 2007, \prl, 75, 115012
\bibitem[Kolb \& Turner (1989)]{kolb89} Kolb, E.W. \& Turner, M.S., \textit{The early Universe}, Addison-Wesley, Redwood City, USA 1989
\bibitem[Lopes et al. (2010)]{lopes10} Lopes, I. \& Silk, J. 2010, Science, 330, 462
\bibitem[Lopes \& Silk(2010)]{lopes10b} Lopes, I., \& Silk, J.\ 2010, \apjl, 722, L95 
\bibitem[Lopes et al. (2011)]{lopes11} Lopes, I., Casanellas, J. \& Eug\'enio, D. 2011, \prd, 83, 63521
\bibitem[Lopes et al.(2014)]{lopes14} Lopes, I., Panci, P.,  \& Silk, J.\ 2014, \apj, 795, 162 
\bibitem[Mao et al. (2013)]{mao13} Mao, Y.Y., Strigari, L.E., Wechsler, R.H., Wu, H.Y. \& Oliver, H. 2013, \apj, 764, 35
\bibitem[Mao et al. (2014)]{mao14} Mao, Y.Y., Strigar, L.E. \& Wechsler, R. H. 2014, \prd, 89, 063513
\bibitem[Morel (1997)]{morel97} Morel, P. 1997, A \& A Supplement Series, 124, 59
\bibitem[Pato et al. (2010)]{pato10} Pato, M., Agertz, O., Bertone, G., Moore, B. \& Teyssier, R. 2010, \prd 82, 023531
\bibitem[Salucci et al. (2010)]{salucci10} Salucci, P., Nesti, F., Gentile, G. \& Martins, C.F. 2010, A\&A, 523, A83 
\bibitem[Serenelli et al. (2009)]{serenelli09} Serenelli, A.M., Basu, S., Ferguson, J.W. \& Asplund, M. 2009, The Astrophysical Journal Letters, 705, L123
\bibitem[Shelton et al. (2015)]{shelton15} Shelton, J., Shapiro, S., \& Fields, B. 2015, \prl, 115, 23
\bibitem[Sheldon et al. (2010)]{sheldon10} Sheldon, C., Dutta, B., \&
    Komatsu, E.  2010, \prd, 82, 9
\bibitem[Silk et al. (1985)]{silk85} Silk, J., Olive, K., Srednicki, M. 1985, \prl, 55, 257 
\bibitem[Sj\"ostrand et al. (2006)]{pythia06} Sj\"ostrand, T., Mrenna, S. \& Skands, P. 2006, J. High Energy Phys. , 05, 026
\bibitem[Sloane et al. (2016)]{sloane16} Sloane, J.D., Buckley, M.R., Brooks, A.M. \& Governato, F. 2016, e-Print: arXiv:1601.05402
\bibitem[Spergel \& Press (1985)]{spergel85} Spergel, D.N. \& Press, W.H. 1985, \apj, 294, 663
\bibitem[Steigman et al. (1978)]{steigman78} Steigman, G., Sarazin,C.L., Quintana, H., Faulkner, J. 1978, Astronomical Journal, 83, 1050
\bibitem[Turck-Chieze \& Lopes (1993)]{turck93} Lopes, I. \& Turck-Chieze, S. 1993, \apj, 408, 347
\bibitem[Turck-Chieze et al. (2004)]{turck04} Turck-Chieze, S., Couvidat, S., Piau, L., Ferguson, J., Lambert, P., Ballot, J., Garcia, R.A. \& Nghiem, P.A.P. 2004, \prl, 93, 211102
\bibitem[Wilkstr\"om \& Edsj\"o (2009)]{wilkstrom09} Wilkstrom, G., Edsjo, J. 2009, JCAP, 04, 009
\end{thebibliography}

\end{document}